\pdfoutput=1

\documentclass[DIV=calc, paper=letter, fontsize=10pt, twocolumn]{scrartcl}	 
\usepackage[english]{babel} 
\usepackage[protrusion=true,expansion=true]{microtype} 
\usepackage{amsmath,amsfonts,amsthm} 
\usepackage{mathtools} 
\usepackage{color}
\usepackage[svgnames]{xcolor} 
\usepackage{booktabs} 
\usepackage{fix-cm}	 

\usepackage{sectsty} 
\allsectionsfont{\usefont{OT1}{phv}{m}{n}} 

\usepackage{fancyhdr} 
\usepackage{lastpage} 

\usepackage[colorlinks=true,urlcolor=black,pagecolor=black,citecolor=black,linkcolor=black]{hyperref}

\usepackage{graphicx}
\usepackage{wrapfig}  
\usepackage{sidecap}  
\usepackage{float}
\usepackage[small,bf]{caption}
\setcapindent{0pt}
\usepackage{wrapfig}
\usepackage{dblfloatfix}
\usepackage{fixltx2e}

\usepackage{lettrine} 

\usepackage{titling} 

\newcommand{\HorRule}{\color{DarkGoldenrod} \rule{\linewidth}{0.5pt}} 

\pretitle{\vspace{-0.75in} \begin{flushleft} \HorRule \fontsize{17}{17} \usefont{OT1}{phv}{b}{n} \color{DarkBlue} \selectfont \vspace{0em}} 

\title{\textrm{Development \& implementation of a PyMOL `\textsf{putty}' representation}} 
\posttitle{\par\end{flushleft}\vskip 0em \vspace{-1.2em}} 

\preauthor{\begin{flushleft} \lineskip -1.2em \usefont{OT1}{phv}{}{} \color{Black}} 

\author{Cameron Mura, } 

\postauthor{\usefont{OT1}{phv}{}{} \color{Black} 
Department of Chemistry \& Biochemistry, UCSD, 
{\large \tt \href{mailto:cmura@mccammon.ucsd.edu}{cmura@mccammon.ucsd.edu}}, 10-Dec-2004 
\vspace{-1.7em}
\par\end{flushleft}\HorRule\vspace{-2.5em}} 

\date{} 

\usepackage{xmpincl}
\include{CC_Attribution-ShareAlike_3.0_Unported}
\usepackage{ccicons}

\newcommand{\mypymol}{\begin{sffamily}PyMOL\end{sffamily}}

\begin{document}

\maketitle 

\thispagestyle{fancy} 

\begin{abstract}\noindent\textbf{Overview} --- The \mypymol~molecular graphics
program has been modified to introduce a new `\textsf{putty}' cartoon
representation, akin to the `\textsf{sausage}'-style representation of the
\textsf{MOLMOL} molecular visualization (MolVis) software package. This document
outlines the development and implementation of the \textsf{putty}
representation.
\end{abstract}

\vspace{-0.75em}
\section*{Background, Motivation}
\vspace{-0.5em}
\mypymol~(\url{http://pymol.org}) is an open-source software package designed
for molecular graphics visualization and modelling.  The \mypymol~code-base is
written in the C programming language, and much of the higher-level,
user-accessible functionality is coded in Python. This software engineering
design endows \mypymol~with a rich Python application programming interface
(API), which end-users can utilize for achieving non-trivial tasks such as
loading and analyzing MD trajectories, parsing and analyzing virus structures,
and so on. In short, \mypymol~is highly \emph{extensible}.

My motivation for introducing `\textsf{sausage}'--style cartoons
stemmed from a desire to represent arbitrary molecular properties
in a cartoon format, such that the diameter of the tubular spline that
traces the backbone of the biopolymer would vary proportionately with 
some property of interest (as encoded in the $B$-factor field of the PDB file). 
The only other molecular graphics program which I am aware of as 
being able to render variable-width cartoons is the \textsf{MOLMOL} 
software from the W\"{u}thrich lab (ETH-Z\"{u}rich); in that case, 
the MolVis software is typically used to represent, say, the 
position-specific RMSD in a bundle of NMR structures.

In the case of MD simulations, one may wish to use such a representation
style to visualize and display the coordinate root-mean-square 
fluctuation (RMSF) along a dynamics trajectory.  Also, 
inclusion of `\textsf{sausage}'-style graphical representations has 
been on \mypymol's ToDo list for some time (personal communication, 
WL DeLano).

\vspace{-0.75em}
\section*{Implementation}
\vspace{-0.5em}
Creation of the new cartoon representation entailed modification of the 
following files in the \mypymol~source code (paths are relative to the base 
of the source tree): 
\begin{verbatim}
  layer1/Extrude.c
  layer1/Extrude.h
  layer1/Rep.h
  layer1/Setting.c
  layer1/Setting.h
  layer2/RepCartoon.c
  modules/pymol/setting.py
  modules/pymol/viewing.py
\end{verbatim}
\noindent
The new rep is termed `\textsf{putty}' so as to avoid conflicts and confusion 
with preexisting `\textsf{sausage}' objects in the \mypymol~source. I had 
initially implemented \textsf{putty} \emph{via} a look-up table approach,
together with variable-width window smoothing of the various vectors and normals 
that govern the geometric and lighting characteristics of the final cartoon in the 
final rendered scene (specular reflections, radiance, radiosity, \emph{etc.}; 
see the `\emph{rendering equation}' and texts on computer graphics rendering for 
more on this, such as Hanrahan's chapter in \cite{Cohen:1993:RRI:154731}). However, 
in numerical benchmarking tests the initial approach of variable window smoothing 
was found to be less efficient than a simple C$\alpha$-based smoothing for 
real-time rendering, so the latter approach was ultimately adopted.

A pictorial overview of the successive stages in \textsf{putty}'s implementation 
can be found at \url{http://muralab.org/~cmura/PyMOL/Sandbox/#putty_work} and in
Figs 1 \& 2 below. For concrete details on \textsf{putty}'s implementation, 
`\texttt{diff}' the codebase just before and just after the stable (v0.98) release, 
paying particular attention to the specific source and header files listed above.

\begin{figure}[H]
  \begin{minipage}[l]{1.0\columnwidth}
    \centering
    \includegraphics[width=0.85\textwidth]{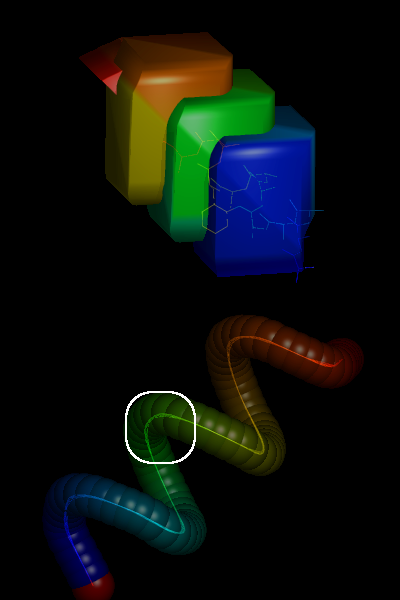}
    \caption{Two very early stages in the implementation of the 
    \textsf{putty} cartoon style. The white-circled region in the bottom
    sub-panel is the region that is magnified in the next figure, revealing the
    key `\emph{extrusion points}' used in this graphical representation.}\label{fig:develop_1b}
  \end{minipage}
\end{figure}

The initial \textsf{putty} code included a provision for anisotropic scaling factors
(\emph{i.e.}, non-circular cross-sections of the rendered cartoon), but this
probably will have to wait for later versions of \mypymol.  The
current \textsf{putty} parameters include a few adjustable/user-input values
(\emph{e.g.}, \texttt{putty\_radius}, \texttt{putty\_quality}), as well as
scaling of tube diameters by statistical properties of the $B$-factors
(\texttt{putty\_mean}, \texttt{putty\_stdev}) such that the default range of
diameters spans $\pm2\sigma$ (\emph{i.e.}, 95\% of the variance, assuming a normal 
distribution). The new \textsf{putty} code has been incorporated into the latest 
CVS versions of \mypymol~and will be part of the next stable release, ca.
early 2005 (v0.98).

\mypymol~does not adhere to the GPL (or any other ``free'' license), so copyright 
interests have been transferred to DeLano Scientific, LLC.

\begin{figure}[H]
  \begin{minipage}[l]{1.0\columnwidth}
    \centering
    \includegraphics[width=0.89\textwidth]{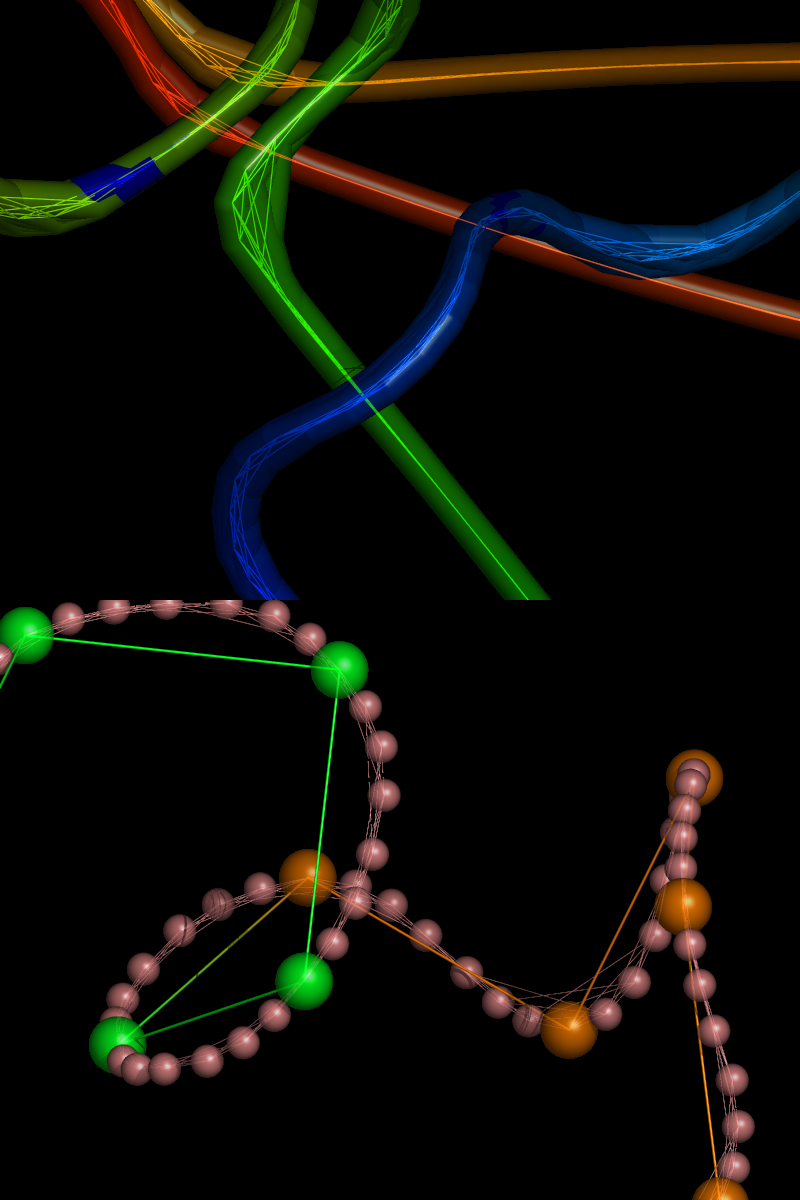}
    \caption{Two later stages (top, bottom) in the implementation of \textsf{putty}. 
    The key `\emph{extrusion points}' (see Fig \ref{fig:develop_1b}) are shown
    as spheres in the bottom panel.}\label{fig:develop_2}
  \end{minipage}
\end{figure}

\section*{Use Cases}
\begin{figure}[H]
  \begin{minipage}[l]{1.0\columnwidth}
    \centering
    \includegraphics[width=0.92\textwidth]{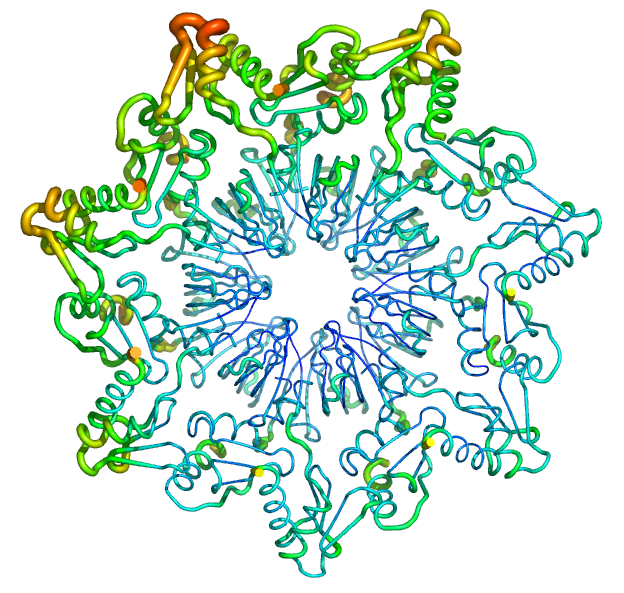}
    \caption{The \emph{Pae} SmAP3 14-mer (PDB 1M5Q; \cite{MuraSmAP3_2003}) is 
    shown rendered as \textsf{putty}. Note that the tubular spline which traces 
    the backbone is also graded in color from lowest (blue) $\rightarrow$ highest 
    (red) $B$-factor values.}\label{fig:smap3}
  \end{minipage}
\end{figure}

\begin{figure}[H]
  \begin{minipage}[l]{1.0\columnwidth}
    \centering
    \includegraphics[width=0.92\textwidth]{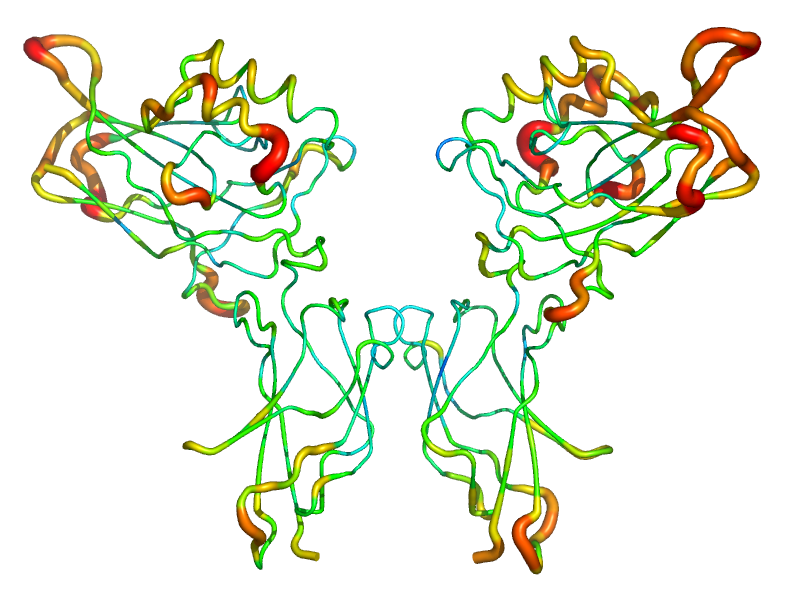}
    \caption{The 3D structure of the `c-Rel' NF-$\kappa$B homodimer \cite{nfkb1:struc}
    is shown rendered as a \textsf{putty} cartoon. As in the previous figure, 
    the tubular spline which traces the backbone is graded in color from 
    lowest (blue) $\rightarrow$ highest (red) $B$-factor values.}\label{fig:nfkb}
  \end{minipage}
\end{figure}

As test cases, the figures shown below demonstrate the application of the
new \textsf{putty}-style cartoon representation to a SmAP3 tetradecamer 
(Fig \ref{fig:smap3}; 1M5Q; \cite{MuraSmAP3_2003}) 
and an NF-$\kappa$B dimer (Fig \ref{fig:nfkb}; 1GJI; \cite{nfkb1:struc}). 
The asymmetric unit of the SmAP3 crystals consists of a 28-mer, so the structure
illustrated in this figure represents only half of the ASU. Not surprisingly, 
residues with the highest $B$-factors in both the SmAP3 and c-Rel/NF-$\kappa$B
crystal structures are located either near the periphery of the oligomer
in the case of SmAP3 (on a face that mediates weak crystal contacts [upper-left];
not the 14-mer$\leftrightarrow$14-mer interface area [lower-right]), or else 
in the irregular/`loopy' regions (distal to the central DNA-binding cleft) in 
the case of the c-Rel transcription factor.

\section*{Acknowledgements}
WL DeLano, creator and chief author of \mypymol, is thanked for answering
questions about the source code and for helping seamlessly integrate the
\textsf{putty} modifications into the final code-base prior to its production
release. This work was done in the McCammon group at UCSD and was partly funded
by a Sloan/DOE postdoctoral fellowship in computational molecular biology. 

\bibliographystyle{plain}
\bibliography{cm5_2004_pymol-putty_v3}

\end{document}